\documentclass[moor]{informs1}              

\usepackage{natbib}
 \NatBibNumeric
 \def\bibfont{\small}%
 \bibpunct[, ]{[}{]}{,}{n}{}{,}%

\usepackage[colorlinks=true,breaklinks=true,bookmarks=true,urlcolor=blue,
     citecolor=blue,linkcolor=blue,bookmarksopen=false,draft=false]{hyperref}

\def\EMAIL#1{\href{mailto:#1}{#1}}

\TheoremsNumberedThrough     

\EquationsNumberedThrough    

\usepackage{graphicx}
\DeclareMathOperator{\Tr}{Tr}

\def\lsim{\mathrel{\rlap{\lower4pt\hbox{\hskip1pt$\sim$}}
    \raise1pt\hbox{$<$}}}                
\def\gsim{\mathrel{\rlap{\lower4pt\hbox{\hskip1pt$\sim$}}
    \raise1pt\hbox{$>$}}}                

\begin{document}





\RUNAUTHOR{Narayan, Saniee, Marbukh}

\RUNTITLE{Congestion Due to Random Walk Routing}

\TITLE{Congestion Due to Random Walk Routing}


\ARTICLEAUTHORS{%
\AUTHOR{Onuttom Narayan}
\AFF{Department of Physics, University of California, \\
1156 High Street, Santa Cruz, CA 95064, 
\EMAIL{onarayan@ucsc.edu}} 
\AUTHOR{Iraj Saniee}
\AFF{Mathematics of Networks Department, Bell Labs, Alcatel-Lucent, \\
600 Mountain Avenue, Murray Hill, NJ 07974, 
\EMAIL{iis@research.bell-labs.com}}
\AUTHOR{Vladimir Marbukh}
\AFF{Applied and Computational Mathematics, National Institute of Standards and Technology, \\
100 Bureau Drive, Gaithersburg, MD 20899, 
\EMAIL{vladimir.marbukh@nist.gov}}
}

\ABSTRACT{
In this paper we derive an analytical expression for the mean load 
at each node of an arbitrary undirected graph for the 
uniform multicommodity flow problem under random walk routing.   
We show the mean load is linearly dependent on
the nodal degree with a common multiplier equal to
the sum of the inverses of the non-zero eigenvalue of the graph Laplacian.
Even though some aspects of the mean load value, such as linear dependence
on the nodal degree, are intuitive and may be derived from
the equilibrium distribution of the random walk on the undirected graph, 
the exact expression for the mean load in terms of the full spectrum of the graph 
has not been known before.  Using the explicit expression for the mean
load, we give asymptotic estimates for the load
on a variety of graphs whose spectral density are well known.
We conclude with numerical computation of the mean load for
other well-known graphs without known spectral densities.
}


\maketitle
%

\section{Introduction}
\label{sec:intro}

The study of network capacity, sometimes referred to as load or congestion, 
is over half a century old, and goes back to the pioneering work of 
Ford and Fulkerson~\cite{ff56} and Shannon~\cite{efs56}
for the single commodity and to early attempts~\cite{hu,lomo,shah-mat,pap,osey} 
for the multicommodity flow solutions of the problem. 
This rather large literature provides a characterization of the load or, more
specifically, the minimal capacity required, in terms of sum of
link capacities needed based on cut values, which in case of the single 
commodity model are both necessary and sufficient and for the multicommodity 
case generally provide necessary conditions.

Single commodity or multicommodity network flow models in communication, 
transportation and numerous other settings typically assume shortest path 
routing.  There are natural settings in which alternative routing
not involving shortest paths may be required.  
For example, it may happen that longer routes are used for
load balancing or, in the case of capacitated 
networks, to avoid network expansion~\cite{min,mag}.
Or the inverse problem may be posed: to determine weights 
so that shortest path routes determined from these weights result in
smallest load across the network~\cite{apple}.  Given the universality of 
the network flow model, there are a vast number of applications of the model,
and the list is too large to enumerate here.

There are few analytical results concerning the multicommodity flow problem
with shortest path routing, in the sense of having a closed form solution 
as a function of a small number of parameters characterizing the network and 
the commodities.  
These include characterization of the maximal load for hyperbolic 
graphs~\cite{ns2,edmond}. In this setting, for a network of $N$ nodes 
one assumes 1 unit of (directed) flow 
between all $N(N-1)$ node pairs, and then asks how the load scales
due to shortest path routing as a function of $N$.
This measure is sometimes referred to as the {\it betweenness centrality},
see~\cite{newman2}.

%

In this paper, we study the near opposite of shortest path routing:
when flows are routed in a uniformly random manner, each flow starting 
from its source and moving at each step randomly to a neighboring
node and only stopping when the destination of the flow is reached. More 
specifically, we consider the case when one unit
of traffic, or a single packet, is injected into the network at every time 
step at each
node $i$ for each possible destination node $j\neq i.$ Thus there
are $N(N-1)$ units of traffic (or packets) injected into the $N$-node network at
every time step. The network is assumed to be connected, i.e. have a single component.
We first demonstrate that a steady-state 
distribution is achieved and then derive an expression for the expected flow, 
or the average number of packets passing through each node, 
in terms of the eigenvalues of the graph Laplacian.  To illustrate the
results more concretely, we estimate the largest mean loads for a set of 
networks whose distribution of Laplacian eigenvalues are known.
We note that similar but not identical measures to the expected load at each node 
have been investigated numerically in the context of node ranking, 
see~\cite{newman1}.

The organization of this paper is as follows.  In Section~\ref{sec:cong},
we obtain an expression for the steady state load at each node in the
network: it is determined by the degree of each node, being equal to 
(up to an additive constant) $N d \sum_1^{N-1}\lambda_\alpha^{-1},$ where
$d$ is the degree of the node and $\{\lambda_0,\lambda_1\ldots\lambda_{N-1}\}$ are 
the eigenvalues of the discrete Laplacian operator on the graph in increasing order
(with $\lambda_0 = 0$).
This is in marked contrast to the result for geodesic routing, where
congestion peaks at the network core.  (For scale-free networks,
there is a very good correlation between the degree of a node and
its betweenness with geodesic routing, and so the distribution of
congestion with geodesic and random routing is related, but this
is not true in general.) Moreover, unlike the case for geodesic
routing, there is no qualitative difference between the congestion
behavior of hyperbolic and non-hyperbolic graphs.  The approach
used in this paper is similar to those in Refs.~\cite{rieger}
and~\cite{dorog}, but the questions we ask are slightly
different. 
Section~\ref{sec:analytical} has observations and analytical
results for a few graph models using the expression derived in
Section~\ref{sec:cong}, and Section~\ref{sec:numerical} has numerical
results for several other graph models.


\section{General results}
\label{sec:cong}
\subsection{Time evolution equations}
As described in the previous section, we consider an undirected 
connected graph $G(N,E)$ with $N$ nodes, in which packets of traffic 
are injected at various nodes
in a deterministic manner
and move towards specified destinations. The dynamics are discrete
time, i.e. packets of traffic move from node to node at time
$t=0,1,2,3\ldots.$ At each time step, exactly $N-1$ packets of unit size are
injected into the graph at each node $k,$ with one packet heading
towards each other node in the graph $l \neq k.$ Thus there are
precisely $N(N-1)$ packets injected into the graph at each time step. Any
packet that is present at node $i$ at time step $t$ and whose
destination is not $i$ moves to one of the nodes adjacent to $i$
at time $t+1.$ For this, one of the $d_i$ nodes adjacent to $i$ is
chosen randomly, with probability 
equal
to $1/d_i$.
However, any packet of traffic that is at its
destination at time $t$ is removed from the network, and is no
longer present at time $t+1.$ Note that a packet that returns to
its source as it moves around randomly continues as it would from
any other node.  The congestion or load at any node at any time
step is 
is a random variable 
equal to the number of packets that are being processed at
that node.
We are interested in the expected value of the number of packets 
at each node.
We expect that in steady state, if and when it exists, 
packets are (injected and) removed from each node
at the same  rate, i.e.  $N-1$ packets per time step.  We seek to
find the steady state load, i.e. the average number of packets, at all the
nodes of the network. 

As a byproduct, we obtain the average time
$\tau$ (or number of steps in its path) that a packet takes to go
from a randomly chosen source node to a randomly chosen destination.
A packet that hops from source to destination in $t$ steps is in
the network for $t$ time steps. (We have assigned one time step
each to the source and destination nodes.) 
The average number of packets
at each node, summed over all the nodes in the graph, is 
therefore the product of the total injection rate $N(N-1)$ and $\tau.$  
\begin{remark}
We shall use $N$ to represent both the set of nodes in the graph
as well as their count $|N|$ without danger of confusion.
Also, we write $k \sim j$ to mean that node $k$ is a neighbor of node $j$,
i.e., $i$ and $j$ are adjacent, 
and $k \nsim j$ when they are not; and refer to the adjacency matrix $(A_{ij})$ 
the Laplacian $(L_{ij})$ and the normalized Laplacian 
$(\cal L_{ij})$ (for $0\leq i,j \leq N$) of the undirected graph $G(N,E)$, 
with their standard definitions:
%
\begin{equation}
\label{eq:min1}
A_{ij} = \left\{
	\begin{array}{rl}
        & 0, \text{          } i = j\\
	& 1, \text{          } i \sim j \\
	& 0, \text{          } i \nsim j
	\end{array} \right.
\\ 
\text{,~~~~~~~~~~~}
L_{ij} = \left\{
	\begin{array}{rl}
	& d_i,  \text{          } \,\,\, i=j \\
	& -1,  \text{          } i \sim j \\
	& 0,  \text{          }\,\,\,\,\, i \nsim j
	\end{array} \right.
\\
\text{,~~~~~~~~~~~}
\cal{L}_{ij} = \left\{
	\begin{array}{rl}
	& 1,  \text{          } \qquad\qquad i=j \\
	& -(d_i d_j)^{-\frac{1}{2}}, \text{          }i \sim j \\
	& 0,  \text{          } \qquad\qquad i \nsim j
	\end{array} \right.
\end{equation}
\end{remark}

\begin{theorem}
\label{existence}
For a connected graph $G(N,E)$ with deterministic injection
rate of one packet at each node destined for each other
node, where each packet is routed uniformly randomly from its
current node to its neighbors until it reaches its destination, 
there exists a unique steady state number of packets at each node.
\end{theorem}

\proof{Proof.}
We first consider the case of the traffic flowing from a single
source node $k$ to a single destination node $l.$ 
Let $X_i^{kl}(t)$ be the random variable representing the
number of packets at node $i$ at time $t$ and $Z_{ji}^{kl}(t+1)$
be the random variable representing the number of packets sent
out of node $j$, a neighbor of $i$, to $i$ at time $t$.  This assumes tacitly that
an outgoing packet from node $j$ that leaves $j$ at time $t$ reaches a neighboring node
$i$ at time $t + 1;$ 
an incoming packet to node $i$ 
from node $j$ that reaches $i$ at time $t$ must leave node $j$ at time $t-1$.
Then the boundary
condition Eq. (\ref{eq0}), and the 
no-escape condition from destination $l$ Eq. (\ref{eq003}), both hold:
\begin{equation}
\label{eq0}
X_i^{kl}(0) = \delta_{ik},  \text{~~~}0\leq i \leq N
\end{equation}
\begin{equation}
\label{eq003}
Z_{li}^{kl}(t)  = 0,  \text{~~~}\forall t \geq 0. 
\end{equation}
Flow balance for outgoing packets implies that for all neighbors $i$ of a node $j \neq l$,
\begin{equation} 
\label{eq008}
X_j^{kl}(t) = \sum_{i\sim j \neq l} Z_{ji}^{kl}(t+1), \text{~~~}1\leq i, j \leq N, 0\leq t.
\end{equation}
which simply states that packets at node $j$ at time $t$ move out
to its incident links at time $t+1$. These same packets arrive at time $t+1$ at adjacent nodes  
\begin{equation} 
\label{eq01}
X_i^{kl}(t+1) = \delta_{ik}+\sum_{l \neq j \sim i} Z_{ji}^{kl}(t+1), \text{~~~}1\leq i, j \leq N, 0\leq t,
\end{equation}
Notice that the first term on the right hand side of Eq.(\ref{eq01}) 
accounts for the fact that
one packet is injected at node $k$ for destination $l$ at each time
step. The second term represents the packets that move to node $i$
at time $t+1$ from adjacent nodes at time $t.$ The sum in this term
excludes the node $l$ because any packet that was at the node $l$
(the destination) at time $t$ is removed from the network and is
no longer present at time $t+1.$

Further, our assumption of
uniformly random routing of packets from each node to its neighbors
implies that for any neighbor $i$ of a node $j \neq l$, 
\begin{equation} 
\label{eq02}
\mathbb{P}\{Z_{ji}^{kl}(t+1)=z\} = \binom {X_j(t)}{z} (\frac{1}{d_j})^z (1-\frac{1} {d_j})^{X_j(t)-z}, \text{~~~} 0 \leq z \leq X_j(t), 0\leq t.
\end{equation}
Taking ensemble expectation of Eqs. (\ref{eq02}) and (\ref{eq01}) 
and using the standard expression for the mean of the
binomial distribution for Eq. (\ref{eq02}), 
we get that for all $0\leq i, j \leq N$
\begin{equation}
\label{eq04}
\mathbb{E} [ Z_{ji}^{kl}(t+1)] = \frac{1}{d_j}\mathbb{E}[X_j^{kl}(t)], \text{~~~} l \neq j \sim i \end{equation}
\begin{equation}
\label{eq03}
\mathbb{E} [ X_i^{kl}(t+1)] = \delta_{ki}+\sum_{l \neq j \sim i} \mathbb{E}[Z_{ji}^{kl}(t+1)] 
\end{equation}
and substituting from Eq. (\ref{eq04}) into (\ref{eq03}), we get
\begin{equation}
\label{eq11}
\mathbb{E}[X_i^{kl}(t + 1)] =
\delta_{ik} + \sum_{l\neq j \sim i} \frac{\mathbb{E}[X_j^{kl}(t)]}{d_j}
\end{equation}
or alternatively stated in terms of the adjacency matrix $A_{ij}$ of the graph,
\begin{equation}
\label{eq12}
\mathbb{E}[X_i^{kl}(t + 1)] = 
\delta_{ik} + \sum_{j\neq l} A_{ij}\frac{\mathbb{E}[X_j^{kl}(t)]}{d_j}.
\end{equation}
Now define $p^{kl}_i(t)= (1 - \delta_{il}) \mathbb{E}[X_j^{kl}(t)].$ In other
words, $p^{kl}_i(t)= \mathbb{E}[X_j^{kl}(t)]$ except for the destination node,
$i=l,$ where $p^{kl}_l = 0.$ The sum in Eq.(\ref{eq12}) can now be
unrestricted for $i\neq l.$ The rate equation for the $p_i$'s is
\begin{equation}
p^{kl}_i(t + 1) = \delta_{ik} + \sum_j A_{ij} \frac{p^{kl}_j(t)}{d_j}
\label{ee2}
\end{equation}
for $i\neq l,$ with the boundary condition $p^{kl}_l(t+1) = 0.$ The
restricted sum in Eq.(\ref{eq11}) has been replaced by an unrestricted
sum in Eq.(\ref{ee2}), but the $l$'th node is now outside the domain
of the equation.  The boundary condition is an example of a Dirichlet
boundary condition, where a function is defined in a region and is specified 
to be zero on the boundary of the region;
in this case, the boundary is the node $l$
and the region is all the other nodes in the graph.

We now show that, under the time evolution of Eq.(\ref{ee2}),
the function $p^{kl}_i(t)$ reaches a $t$-independent unique steady state.
Let $p^{kl(1)}_i(0)$ and $p^{kl(2)}_i(0)$ be two initial configurations
at $t=0,$ that evolve according to Eq.(\ref{ee2}). Define $q^{kl}_i(t)$
to be equal to $[p^{kl(1)}_i(t) - p^{kl(2)}_i(t)]/\sqrt{d_i}.$ Then $q^{kl}$
satisfies 
\begin{equation}
q^{kl}_i(t + 1) = \sum_j A_{ij} \frac{q^{kl}_j(t)}{\sqrt{d_j d_i}}
\label{pf}
\end{equation}
with the Dirichlet boundary condition at $i=l.$ This is equivalent
to $q^{kl}_i(t + 1) = \sum_j (\delta_{ij} - {\cal L}_{ij}) q^{kl}_j(t),$
where ${\cal L}$ is the normalized Laplacian. Since ${\cal L}$ is a real 
symmetric matrix, it has a complete set of eigenfunctions. The eigenvalues
are all in the interval $0\leq \lambda \leq 2,$ with an eigenvalue at $\lambda = 0$ 
iff one can construct a function $f$ on the graph for which $f_i = f_j$ for all nodes 
$(i, j)$, and an eigenvalue at $\lambda = 2$ iff one can 
construct $f$ such that $f_i = - f_j$ whenever $j\sim i$~\cite{fan}. With 
Dirichlet boundary conditions, since $f=0$ on the boundary nodes, both of these are impossible,
and therefore $0 < \lambda < 2.$ Thus the operator $I - {\cal L}$ (with Dirichlet boundary
conditions) is a contraction. 
Therefore $q^{kl}(t\rightarrow\infty)\rightarrow 0,$ and 
as $t\rightarrow\infty$ all initial configurations tend to the same $t$-independent 
steady state configuration.
\qed
\endproof

\subsection{Steady state solution}
In this section, we solve the fixed point of the time evolution 
equation~(\ref{ee2})
with Dirichlet boundary condition as introduced in the proof of Theorem~(\ref{existence}).
As before, $\{\lambda_{\alpha},\alpha<N\}$ represent the eigenvalues of the graph
Laplacian.
\begin{theorem}
\label{solution}
For a connected graph $G(N,E)$ with deterministic injection
rate of $(N-1)$ packets at each node destined for every other
node, where each packet is routed uniformly randomly from its
current node to its neighbors until it reaches its destination, 
the unique steady state number of packets at each node $j$
is given by $\Lambda_j$ where
\begin{equation}
\Lambda_j = (N - 1) + N d_j \sum_{\alpha \neq 0}\frac{1}{\lambda_\alpha}.
\label{qw-1}
\end{equation} 
\end{theorem}
\proof{Proof.}
In steady state, we know
that the load flowing into the node $l$ at any time step must be
equal to the load injected into the node $k,$ i.e. unity. Therefore
$\sum A_{lj} p^{kl}_j/d_j = 1,$ and we can extend Eq.(\ref{ee2}) as
\begin{equation}
p^{kl}_i= \delta_{ik} - \delta_{il} + \sum_j A_{ij} \frac{p^{kl}_j}{d_j}
\label{ee3}
\end{equation}
for all $i,$ with the additional condition $p^{kl}_l = 0.$ It may
seem that we have gained nothing by restricting our analysis to the
steady state configuration, since we still have to impose Dirichlet
boundary conditions at the $l$'th node. However, as we shall see
immediately, the solution to Eq.(\ref{ee3}) can easily be found in
terms of the eigenvectors of the Laplacian without the Dirichlet
boundary condition, i.e. independent of $k$ and $l.$

In order to convert Eq.(\ref{ee3}) to a Hermitean eigenvalue problem,
we define $p^{kl}_j = d_j r^{kl}_j$ and $L_{ij} = d_j\delta_{ij} -
A_{ij}.$ Then
\begin{equation}
\sum_j L_{ij} r^{kl}_j = \delta_{ik} - \delta_{il}
\label{ee4}
\end{equation}
with $r^{kl}_l = 0.$ Here $(L_{ij})$ is the Laplacian for the graph.
Since $(L_{ij})$ is a real symmetric matrix, it has a complete set
of real eigenvalues $\lambda_\alpha$ and real orthonormal eigenvectors
$\xi^\alpha$ for $\alpha = 0, 1, 2\ldots N - 1.$ Using the standard
properties of the Laplacian, all the eigenvalues are
non-negative, and since the graph has been assumed to have one
component, there is only one zero eigenvalue $\lambda_0$ with
eigenvector $\xi^0 = (1, 1, 1,\ldots 1)/\sqrt N.$ The denominator
ensures that the normalization condition $\sum_i \xi^0_i \xi^0_i =
1$ is satisfied.

We define 
\begin{equation}
\pi_{kl}^\alpha = \sum_i \xi^\alpha_i (\delta_{ik} - \delta_{il}) 
   = \xi^\alpha_k - \xi^\alpha_l
\end{equation}
which is the projection of the right hand side of Eq.(\ref{ee4})
on to the $\alpha$'th eigenvector. Note that $\pi_{kl}^0 = 0.$
With this definition,
\begin{equation}
r^{kl}_j = \sum_{\alpha = 1}^{N-1} \frac{\pi_{kl}^\alpha}{\lambda_\alpha}\xi^\alpha_j 
+ c^{kl} \xi^0_j,
\end{equation}
where $c^{kl}$ has to be chosen to make $r^{kl}_l$ equal to zero. Since $\xi^0_j$ is
independent of $j,$ the condition $r^{kl}_l=0$ yields
\begin{equation}
r^{kl}_j = \sum_{\alpha = 1}^{N-1} \frac{
\xi^\alpha_k - \xi^\alpha_l}{\lambda_\alpha}\xi^\alpha_j  
- \sum_{\alpha = 1}^{N-1} \frac{1}{\lambda_\alpha}
[\xi^\alpha_k\xi^\alpha_l - (\xi^\alpha_l)^2].
\label{qj}
\end{equation}

Averaging over all the random paths taken by the traffic packets,
the steady state load at any node $j\neq l$ is $p^{kl}_j = r^{kl}_j
d_j.$ For the $l$'th node, the load is $\mathbb{E}[X_l^{kl}] \neq
p^{kl}_l,$ since we defined $p_l^{kl}$ to be zero. However, in
steady state we know that the traffic flowing out of node $l$ at
any time step is unity, and this is equal to the entire load $\mathbb{E}[X_l^{kl}]$
at that time step. Therefore, in steady state, the load
at the $j$'th node is equal to $\Lambda_j^{kl} = d_j r_j^{kl} +
\delta_{jl}.$ Note that a unit of load from $k$ to $l$ is counted
at all the nodes it passes through, as well as the source and
destination nodes. Depending on how traffic is actually processed
by the network, it may be appropriate to change the weightage given
to the source and destination nodes.

Summing over all source destination pairs, the total steady state
load at the $j$'th node is
\begin{equation}
\Lambda_j = d_j \sum_l \sum_{k\neq l} r_j^{kl} + N - 1.
\end{equation}
Since the first term on the right hand side of Eq.(\ref{qj}) is
antisymmetric in $k$ and $l,$ only the second term contributes to
$\sum_l\sum_{k\neq l} r_j^{kl}.$ In the second term, we can replace
the sum $\sum_{k\neq l}$ with an unrestricted sum over $k,$ so that
\begin{eqnarray} \Lambda_j 
  &=& (N -1) + d_j \sum_{\alpha = 1}^{N-1} \frac{1}{\lambda_\alpha}
\Big[N \sum_l (\xi^\alpha_l)^2 - (\sum_l \xi^\alpha_l)^2 \Big] \nonumber\\
  &=& (N - 1) + N d_j \sum_{\alpha \neq 0}\frac{1}{\lambda_\alpha}.
\label{qw2}
\end{eqnarray}
The load $\Lambda_j$ at any node $j$ is linearly dependent on the
degree $d_j$ of the node. Unlike the case when traffic between any
source and destination flows along the geodesic path connecting
them, there is no concept of a network core.
\qed
\endproof

\begin{remark}
The result $\Lambda_j - (N - 1) \propto d_j$ can be obtained
directly. An outline of the proof is as follows.
The traffic from node $k$ to node $l$ can be represented
as a stream of random walkers that diffuse through the network at
discrete time steps. At every time step in addition to the diffusive
dynamics, a walker is introduced at node $k,$ and all the walkers
at node $l$ are removed. Comparing with Eq.(\ref{ee2}), the 
expected number
of random walkers at node $j$ at time $t$ is equal to $p^{kl}_j(t).$
If the random walks corresponding to all source destination pairs
take place simultaneously, with each walker labelled with an index
corresponding to its destination, we have random walkers with $N$
different labels moving through the network. In addition to the
random walk dynamics, walkers are created and destroyed at their
sources and destinations respectively. In steady state, the number
of walkers created and destroyed at any time step are equal to $N-1$
at each node, but they have different labels. If we ignore the
labels on the random walkers, the creation and destruction of random
walkers can be ignored. The steady state solution
for $\sum_k \sum_l p^{kl}_j(t)$ is proportional to the steady state
solution for a diffusion process on the graph with no sources or
sinks.  It is easy to verify that, in this steady state, the number
of random walkers at any node is proportional to the degree of the
node. Although this tells us that $[\Lambda_j - (N - 1)]/d_j$ is
a constant, independent of $j,$ it does not tell us that this
constant is equal to $N\sum_{\alpha\neq 0} 1/\lambda_\alpha.$
\end{remark}

\begin{remark}
If instead of using the Laplacian, $L$, of the graph, we had
used the normalized Laplacian, ${\cal L}$, the entire proof
would have proceeded as presented except that equation~(\ref{qw2})
would have read as follows
\begin{eqnarray} 
\Lambda_j 
  &=& (N -1) + d_j \sum_{\alpha = 1}^{N-1} \frac{1}{\nu_\alpha}
\Big[N \sum_l (\frac{\zeta^\alpha_l}{\sqrt{d_l}})^2 - (\sum_l \frac{\xi^\alpha_l}{\sqrt{d_l}})^2 \Big] \nonumber\\
  &=& (N - 1) + N d_j \sum_{\alpha \neq 0}\frac{1}{\nu_\alpha} Var (\frac{\zeta^\alpha_l}{\sqrt{d_l}}).
\label{qw2-0}
\end{eqnarray}
where $0 \leq \nu_0, ... \leq \nu_{N-1} \leq 2$ are the eigenvalues 
and $\{\boldsymbol\zeta_\alpha \}$ are the corresponding orthonormal
eigenvectors of $\cal L$ and 
$0 \leq \lambda_0, ... \leq \lambda_{N-1} \leq 2$ 
and $\{\boldsymbol\xi_\alpha \}$ are the eigenvalues and eigenvectors
of $L$.
We note that expressions involving terms similar to the right-hand 
side of equation (\ref{qw2-0}) were obtained in \cite{Lov} in
the context of hitting times of Markov chains, and it may be possile
to obtain simpler expressions there by using the Laplacian, as 
we did above.
Equations(\ref{qw2}) and (\ref{qw2-0}) give an interesting relationship
between the spectra of the Laplacian and
those of the normalized Laplacian for an arbitrary graph 
which we had not come across before.
\end{remark}
 
\begin{remark}
So far we have dealt with connected undirected graphs.  We point out that
when the graph is directed, then assuming that steady state distribution
is achieved, Remark~2 implies that the expected load $\Lambda_j = N-1 + C \pi_j$ 
where $C$ is some constant independent of the node and
$(\pi_j)$ is the principal eigenvector of the random walk matrix for 
the directed graph, which for undirected graphs is equal to $(d_j)$.

\end{remark}

\begin{remark}
We observe that the proofs of both theorems carry through essentially 
unchanged if we replace
the deterministic arrival of one packet at
each source node for each destination node at each time step with a Poisson arrival process with a mean 
of one packet arrival per node per
unit time for each destination node. The same is true if we replace  
the uniform random routing from each
node to its neighbors with a more general value $w_{jk}/w_j$ with
$w_j=\sum_{l \sim j} w_{jl}$ for the probability
of moving from a node $j$ to any of its neighbors $k,$ 
so long as $w_{jk} = w_{kj} \neq 0.$ 
However, the normalized Laplacian $({\cal L}_{jk})$ and its
eigenvalues $\{\lambda_{\alpha},\alpha<N\}$
in Theorem~(\ref{solution}) are now replaced by 
$({\cal L}^{\textbf{w}}_{jk})$ and its eigenvalues $\{\lambda^{\textbf{w}}_{\alpha},\alpha<N\}$ 
where $({\cal L}^{\textbf{w}}_{jk})$ is now the 
weighted normalized Laplacian~\citep{fan}, 
defined analogously as 
${\cal L}^{\textbf{w}}_{jk} = \delta_{jk}-(1-\delta_{jk})w_{jk}/\sqrt{w_j w_k}$ 
instead of 
${\cal L}_{jk} = \delta_{jk}-(1-\delta_{jk})/\sqrt{d_j d_k}$, see~(\ref{eq:min1}) 
in Remark 1.
\end{remark}

\subsection{Discussion}
In the large-$N$ limit, the spectral density of the Laplacian
$\sum_\alpha \delta(\lambda - \lambda_\alpha)$ tends to $N\rho(\lambda)$
where $\rho(\lambda)$ is smooth. If $\rho(\lambda\rightarrow 0) =
0,$ we have 
\begin{equation}
N \sum_{\alpha\neq 0} \frac{1}{\lambda_\alpha} \rightarrow N^2 
\int \frac{\rho(\lambda)}{\lambda} d\lambda\sim N^2
\end{equation}
for large $N.$ The simplest example of this is when the graph
Laplacian has a spectral gap in the large $N$ limit. A more subtle
case is the Erd\"os-R\'enyi model~\cite{e-r}, where the 
spectral density is empirically
found~\cite{tucci} to be close to that of a infinite regular tree
whose nodes all have the same degree as the average degree of the
Erd\"os-R\'enyi graph. Even though the infinite tree has a spectral gap, 
the corresponding Erd\"os-R\'enyi spectral density has a narrow tail extending
down to $\lambda = 0,$ so that there is no spectral gap~\cite{dorog}. 
However, in the next section of this paper, we find
numerically that $N\sum_\alpha \lambda_\alpha^{-1}\sim N^2$ for
Erd\"os-R\'enyi graphs, presumably because the density in the tail as $\lambda\rightarrow 0$
is $\rho(\lambda\rightarrow 0) = 0.$
The same result is also
shown numerically for scale-free graphs.

If $\rho(\lambda\rightarrow 0)$ is not zero, $N \sum_{\alpha\neq
0} 1/\lambda_\alpha$ diverges faster than $\sim N^2$ for large $N.$
If $\rho(\lambda\rightarrow 0)$ is finite, the spectral gap for
large but finite $N$ is proportional to $1/N.$ Then $N^2 \int
\rho(\lambda)/\lambda d\lambda$ diverges as $-N^2\ln\lambda_{min}\sim
N^2\ln N.$ This is the case for the square lattice and, as is shown
in the next section of the paper, a finite regular tree.  For hyperbolic
grids $\mathbb{H}_{p,q}$ (where $p$ and $q$ are positive integers satisfying 
$(p-2)(q-2)>4$), which are infinite regular
planar graphs with constant degree $q$ and $p$-sided polygons as
faces, we show numerically in the next section of this paper that
$N\sum \lambda_\alpha^{-1} \sim N^2\ln N.$ 

The maximum congestion in the network is, up to an additive constant,
equal to the product of $N\sum 1/\lambda_\alpha$ and $d_{max}.$ The
large-$N$ dependence of the latter depends on the degree distribution,
e.g growing as $\sim \ln N$ for Erd\"os-R\'enyi graphs and as a power of 
$N$ for scale-free networks.

The average time $\tau$ that a packet spends in the network is
obtained from the equation $N(N-1) \tau = \sum_j \Lambda_j,$  from
which
\begin{equation}
\tau = \frac{\sum_j d_j}{N - 1} \sum_{\alpha = 1}^{N-1} \frac{1}{\lambda_\alpha}  + 1
\rightarrow \overline d  \sum_{\alpha = 1}^{N-1} \frac{1}{\lambda_\alpha}
\label{tau}
\end{equation}
in the large $N$ limit,
where $\overline d$ is the average degree of nodes in the graph.
If $\sum\lambda_\alpha^{-1} \sim N,$ the average sojourn time in
the graph is $O(N).$ To express this in terms of the diameter of
the graph instead of the number of nodes, we have to know how the
diameter grows as $N$ is increased; for small world graphs, $\tau$
grows exponentially as the diameter of the graph is increased. 
Exponential growth implies that a shortest-path walk starting
at a site $k$ and aimed at site $l$ can reach destination $l$
exponentially faster on average than the random walk.

\section{Analytical results for models}
\label{sec:analytical}
\subsection{Hypercubic lattices}
\begin{figure}
\begin{center}
\includegraphics[width=1.6in]{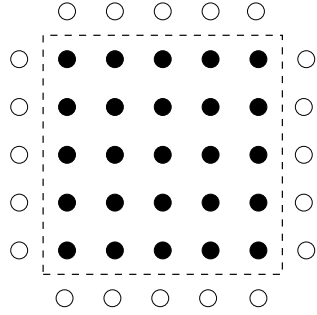}
\caption{A square graph with the nodes arranged in a square lattice.
The black nodes are in the graph. The white nodes are an extra layer
of fictitious nodes that are introduced in order to obtain the
eigenfunctions of the Laplacian operator on the graph.}
\label{fig:square_lat}
\end{center}
\end{figure}
Suppose the graph is a square lattice, with nodes labeled as $\{x_1,
x_2\}$ with $1 \leq x_1, x_2 \leq w.$ Two nodes ${\bf x}$ and ${\bf
y}$ are connected if $\sum_\alpha |x_\alpha - y_\alpha| = 1.$ All
nodes in the interior of the square have four nearest neighbors,
but the nodes on the perimeter have three and the nodes at the
corners have two. The action of the Laplacian
on a node is $L r_{\bf x} = d_{\bf x} r_{\bf x} - \sum_{\bf y}
r_{\bf y},$ where $d_{\bf x}$ is the degree of the node ${\bf x}$
and the sum over ${\bf y}$ is restricted to the nearest neighbors
of ${\bf x}.$ As shown in Figure \ref{fig:square_lat}, if one adds a
layer of new nodes around the perimeter of the square, with $r_{\bf
y}$ at a perimeter node ${\bf y}$ equal to $r_{\bf x}$ at its
adjacent new node, the action of the Laplacian is the same
at all the nodes in the lattice. This corresponds to imposing
discrete Neumann boundary conditions on the lattice. (As compared to Dirichlet 
boundary conditions that have been discussed earlier, Neumann boundary conditions 
for a function defined in a region require that the normal component of the gradient 
of the function should be zero everywhere on the boundary.) 
The eigenfunctions
of the (discrete) Laplacian are then of the form
\begin{equation}
\xi_{m,n}({\bf x}) \propto 
\cos \left[m\pi (x_1 - {1\over 2})/w\right] \cos\left[n\pi (x_2 - {1\over 2})/w\right]
\end{equation}
with eigenvalues 
\begin{equation}
\lambda_{m,n} = 4 - 2 \cos(m\pi/w)) - 2\cos(n\pi/w)
\end{equation}
where $m, n = 0, 1, 2\ldots (w - 1).$ 

\begin{figure}
\begin{center}
\includegraphics[width=2.0in]{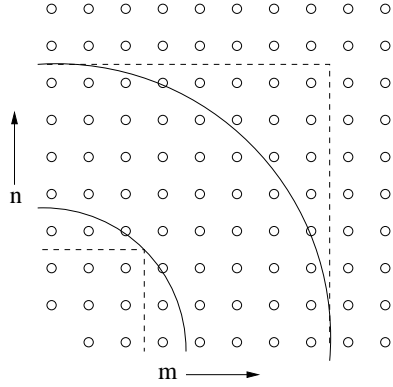}
\caption{Division of the summation in $\sum 1/\lambda_{mn}$ into
three parts. (The point $m=n=0$ is excluded from the sum.) Outside
the inner dashed line, the sum can be replaced by an integral;
inside the outer dashed line, the $\lambda_{mn}$ is approximately
$(m^2 + n^2) \pi^2/w^2.$ The middle part of the summation is all
the $(m,n)$ values that are inside the annulus quadrant.}
\label{fig:range}
\end{center}
\end{figure}

When $w$ is large, the summation in $\sum 1/\lambda_\alpha$ can be
divided into three parts. In the middle part, either $m$ or $n$ is
greater than $N_1,$ where $N_1$ is sufficiently large that the
summation in $\sum 1/\lambda_\alpha$ can be replaced by an integral.
At the same time, $k_x = m/w$ and $k_y = n/w$ are less than $\delta,$
where $\delta$ is sufficiently small that $\lambda_{m,n}$ is
approximately $(m^2 + n^2) \pi^2/w^2.$ The error in these approximations
can be made as small as desired by decreasing $\delta$ and increasing
$N_1.$ For any choice of $(\delta, N_1)$ it is possible to make $w$
sufficiently large that $w\delta >> N_1.$ As shown in
Figure~\ref{fig:range}, for convenience we further reduce the middle
region to be a circular annulus in the $m-n$ plane. In terms of
$(k_x, k_y)$ the inner and outer radii of the annulus are $ N_1
\sqrt 2/w$ and $\delta.$
Then 
\begin{equation}
\begin{split}
\sum_{\alpha\neq 0} \frac{1}{\lambda_\alpha}\rightarrow
&w^2
\int_{|k| > \delta}^{k_x, k_y = 1}
\frac{dk_x dk_y}{4 - 2 \cos \pi k_x - 2 \cos \pi k_y} + \\
&w^2\int_{|k| = N_1\sqrt 2/w}^{\delta} \frac{dk_x dk_y}{\pi^2 (k_x^2 + k_y^2)} +
\sum_{|m|^2 + |n^2| \leq 2 N_1^2} \frac{w^2}{(m^2 + n^2) \pi^2}.
\end{split}
\end{equation}
The first and the third terms on the right hand side are clearly
proportional to $w^2.$ The second term is equal to $w^2 \int (2 \pi
k dk)/k^2 = 2\pi w^2 \ln {w\delta/N_1\sqrt 2}.$ Thus for large $w$,
the leading term in the summation is $2\pi w^2 \ln w = \pi N\ln N,$
since the number of nodes in the graph is $N = w^2.$

It is straightforward to generalize this approach to a $d$-dimensional hypercubic lattice.
For $d > 2,$ the first and second parts of the summation are proportional to $w^d = N,$
while the third part is proportional to $w^2,$ which can be neglected in comparison. For
$d=1,$ the second and third parts are proportional to $w^2 = N^2$ while the first part is
proportional to $w=N$ which can be neglected in comparison. Thus 

\begin{align}
\sum_{\alpha\neq 0} \frac{1}{\lambda_\alpha} \sim  
\begin{cases}
N, &d > 2\nonumber\\
N\ln N, &d = 2 \nonumber\\
N^{2/d}, &d < 2.
\end{cases}
\end{align}

                                           
\subsection{Regular trees}
In this subsection, we calculate $\sum_{\alpha\neq 0} \lambda_\alpha^{-1}$
for a regular tree. We show that the spectrum of a finite tree near
$\lambda = 0$ has a large-$N$ form that results in $\sum_{\alpha\neq
0} \lambda_\alpha^{-1}\sim N\ln N$ for large $N.$ Thus as a function
of the number of levels $h$ in the tree, the average sojourn
time of a random walker in the graph is $\sim h \exp[a h]$ for large
$h.$

\begin{figure}
\begin{center}
\includegraphics[width=3.5in]{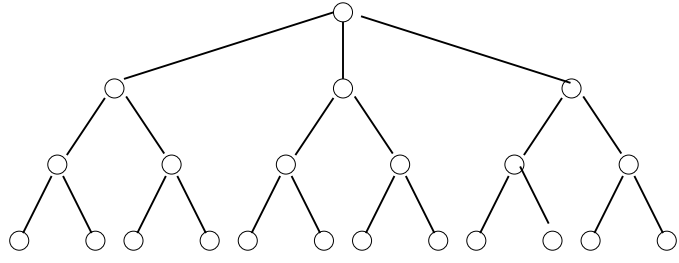}
\caption{Regular tree with $q=3$ nearest neighbors of each node (except the leaf nodes)
and $h=3$ levels below the root node.}
\label{fig:tree}
\end{center}
\end{figure}
The graph is generated from a tree that extends $h$ levels from the
root node, with each node except the leaf nodes at level $h$ having
degree $q.$ The eigenvalues of the Laplacian fall in two categories.

First, we consider eigenvectors whose value is the same at all the
nodes at the same level, i.e. they are azimuthally symmetric. The
eigenvalue equation for $L_{ij}$ reduces to
\begin{equation}
 q \xi_i -  \xi_{i-1} - (q-1) \xi_{i+1} = \lambda \xi_i.
\label{recurs}
\end{equation}
For the root node, which has $q$ daughters, the equation is $q \xi_0
- q \xi_1 = \lambda \xi_0,$ while for the leaf nodes the equation
is $\xi_h - \xi_{h-1} = \lambda \xi_h.$ These nodes also satisfy
Eq.(\ref{recurs}) if we impose the boundary conditions $\xi_{-1} =
\xi_1$ and $\xi_{h+1} = \xi_h.$ The eigenvalues of the Laplacian
are now those of a $h+1\times h + 1$ tridiagonal matrix acting on
the column vector $[\xi_0, \xi_1,\ldots \xi_h],$ instead of a
$N\times N$ matrix:
\begin{equation}
\left(
{\begin{array}{cccccc}
         q & -q & 0 & 0 & 0 & \ldots\\
       -1 & q & -(q - 1) & 0 & 0 & \ldots\\
         0  & -1 & q & -(q - 1) & 0 & \ldots \\
	\vdots &  & \vdots & & \vdots & \\
               & \ldots & 0  & -1 & q & -(q - 1)\\
               & \ldots & 0  &  0   & -1 & 1
               \end{array}}\right).
\end{equation} 
There are $h$ eigenvectors of this matrix excluding the trivial one
where all the $\xi_i$'s are the same and $\lambda = 0.$ They are
obtained by solving the recursion relation Eq.(\ref{recurs}). The
general solution to this equation is $\xi_i = A \rho_1^i + B \rho_2^i$
with $\rho_{1,2}$ the roots of the equation $(q - 1)\rho^2 + 1 +
(\lambda - q) \rho = 0.$ This has solutions $\rho_{1,2} = e^{\pm
i\alpha}/\sqrt{q-1},$ with 
\begin{equation}
\lambda = q - 2 \sqrt{q - 1} \cos\alpha.
\label{ev}
\end{equation}
Since $\lambda$ is real, either the real or the imaginary part of
$\alpha$ must be zero. The boundary conditions are
\begin{eqnarray}
A (\rho_1 - 1/\rho_1) + B(\rho_2 - 1/\rho_2) &=& 0\nonumber\\
A \rho_1^h (\rho_1 - 1) + B \rho_2^h(\rho_2 - 1) &=& 0.
\label{bc1}
\end{eqnarray}
If $\lambda\neq 0,$ $\rho_{1,2}\neq 1$ and $\rho_1^{h+1} (1 + \rho_2)
= \rho_2^{h+1}(1 + \rho_1),$ which is equivalent to $\sqrt{q-1}\sin
(h + 1)\alpha = -\sin h\alpha,$ i.e.
\begin{equation}
\frac{\sqrt{q-1}\sin\alpha}{1 + \sqrt{q-1}\cos\alpha} = -\tan h\alpha.
\end{equation}
There are no solutions to this equation with imaginary $\alpha.$
Therefore all the $h$ non-trivial eigenvalues correspond to real
$\alpha,$ i.e. from Eq.(\ref{ev})
\begin{equation}
     q + 2 \sqrt{q-1} > \lambda > q - 2 \sqrt{q-1}.
\end{equation}
Thus there is a spectral gap between the lowest non-zero eigenvalue
and $\lambda = 0.$ The contribution of these eigenvalues to
$\sum_\alpha 1/\lambda_\alpha$ is $O(h),$ i.e. $O(\ln N).$ (If the
root node has $q-1$ daughters instead of $q,$ the boundary condition
at the root is $\xi_0 = \xi_1,$ which results in the first equation
in Eqs.(\ref{bc1}) being replaced with $A(\rho_1 - 1) + B (\rho_2
- 1) = 0.$ With the second equation, this immediately yields
$\rho_1^h = \rho_2^h,$ i.e.  $\alpha = m \pi/h$ with integer $m,$
and hence a spectral gap.)

Second, we consider eigenvectors which are only non-zero at two
daughters of a node at the $k$'th level and descendants thereof,
with $ h > k \geq 0.$ The eigenvector must be equal and opposite
at the two daughters at the $k+1$'th level in order that the
eigenvalue equation for their parent should be satisfied. Thereafter,
the eigenvector must be azimuthally symmetric within the sectors
descending from either daughter. The recursion relation inside
either of the two sectors is then the same as Eq.(\ref{recurs}),
but the first boundary condition is replaced by $\xi_k = 0.$ There
are $h - k$ eigenvalues for any $k,$ each with degeneracy $N_{k+1}
- N_k,$ where $N_k$ is the number of nodes at the $k$'th level.
Adding up the eigenvectors of both types, we have
\begin{equation}
h + 1 + \sum_{k=0}^{h - 1} [N_{k+1} - N_k] (h - k) = 
  h + 1 + \sum_1^h N_k - h N_0 = N
\end{equation}
and thus we have accounted for all the eigenvectors.

For these second type of eigenvectors, the eigenvalues are those
of the $h - k\times h - k$ matrix $S_k$ that, acting on the
column vector $[x_{k+1}\ldots x_h]$ corresponding to a sector,
codifies the recursion relation and boundary conditions: 
\begin{equation}
S_k = \left(
{\begin{array}{ccccc}
         q & -(q-1) & 0 & 0 & \ldots\\
       -1 & q & -(q - 1) & 0 & \ldots\\
	& \vdots &  & \vdots &  \\
               & \ldots & -1 & q & -(q - 1)\\
               & \ldots &  0   & -1 & 1
               \end{array}}\right).
\end{equation} 
Including degeneracies, the
contribution of these eigenvectors to $\sum_\alpha 1/\lambda_\alpha$
is 
\begin{equation}
\sum_{k=0}^{h-1} \{N_{k + 1} - N_k\} \Tr[S_k^{-1}].
\label{type2}
\end{equation}
To find $\Tr[S_k^{-1}],$ we calculate the cofactors and
determinant of $S_k$ using the tridiagonal form of $S_k.$,
Let $r_n$ and $s_n$ be the determinants of of the top left and
bottom right $n\times n$ submatrices of $S_k$ respectively;
these are independent of $k.$ Then
\begin{eqnarray}
r_n &=& q r_{n-1} - (q-1) r_{n-2}; \qquad r_1 = q, r_0 = 1\nonumber\\
s_n &=& q s_{n-1} - (q-1) s_{n-2}; \qquad s_1 = 1, s_0 = 1
\end{eqnarray}
with the $i$'th cofactor of $S_k$ equal to $r_{i-1} s_{h-k-i}$ and
$\det [S_k] = s_{h-k}.$ The solutions to the recursion equations
for $r_n$ and $s_n$ are
\begin{eqnarray}
r_n &=& \frac{(q - 1)^{n+1} - 1}{q-2}\nonumber\\
s_n &=& 1.
\end{eqnarray}
From these, it is straightforward to obtain
\begin{equation}
\Tr[S_k^{-1}] = 
  \frac{1}{s_{h-k}} \sum_{n = 0}^{h - k - 1} r_n s_{h - k - 1 - n}
= \frac{1}{(q - 2)^2} [(q - 1)^{h - k + 1} - (q - 1) - (h - k) (q - 2)].
\end{equation}
Since $N_k = q (q - 1)^{k-1}$ for $k > 0$ and $N_0
= 1,$ one can perform the summation in Eq.(\ref{type2}). For
$h\rightarrow\infty,$ the expression simplifies to leading order
as
\begin{equation}
N \left[h \{1 + O(1/N)\}  
+ \left\{\frac{2 - 2 q^2 + 2 q^3 -q^4}{2 q (q - 2) (q - 1)^2} + O(1/N)\right\}\right ]
   \rightarrow N h = O(N\ln N).
\end{equation}

\subsection{Numerical results}
\label{sec:numerical}
Except for a few network models discussed in the previous
section, one has to compute the large $N$ behavior
of load and thus the sum $N\sum \lambda_\alpha^{-1}$ numerically. 
In this section, we show the results for a few cases involving
prototypical graphs including the Erd\"os-R\'enyi random graphs
in various regimes, the Barab\'asi-Albert model of preferential 
attachment~\cite{barabasi,krapi}, and hyperbolic grids.

Because of its zero eigenvalue, the matrix $L$ is not invertible.
We define the matrix $M = L + P,$ where $P_{ij} = 1/N.$ Then $P$ is
a projection operator: $P\sum_\alpha c_\alpha \xi^\alpha = c_0\xi^0.$
Therefore
\begin{equation}
M\sum_\alpha c_\alpha \xi^\alpha = \sum_\alpha (\lambda_\alpha +\delta_{\alpha 0}) c_\alpha \xi^\alpha.
\end{equation}
Therefore $M$ is an invertible matrix, with $\Tr[M^{-1}] = \sum_\alpha
(\lambda_\alpha + \delta_{\alpha 0})^{-1},$ which is equal to $1 +
\sum_{\alpha\neq 0} \lambda_\alpha^{-1}.$ We have to numerically evaluate
$\Tr[M^{-1}] - 1.$

\begin{figure}
\begin{center}
\includegraphics[width=3.5in]{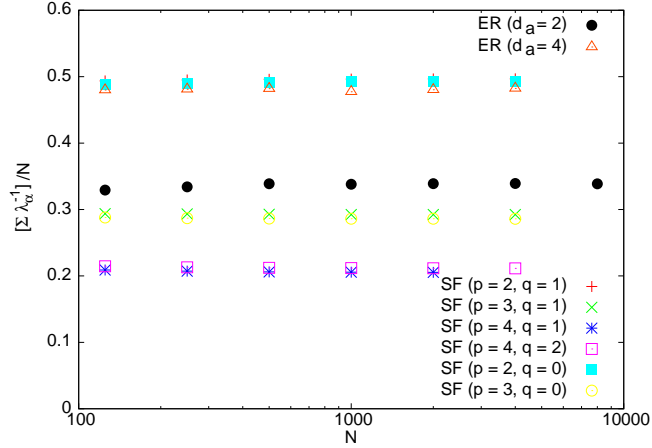}
\caption{Plot of $[\sum_{\alpha\neq 0} \lambda^{-1}_\alpha]/N$
versus $N$ for the Erd\"os-R\'enyi model with average nodal degree of
2 and 4. (For the first of these, the vertical axis is scaled by a 
factor of 0.25 to fit in the figure.) Also shown are the results for scale free networks, where
each node is born with $p$ edges that link it to preexisting nodes,
and the probability of linking to a preexisting node is proportional
to its degree with an offset of $q$; the results for various values
of $(p,q)$ are shown. The curves are flat for all the cases,
demonstrating that $N\sum \lambda_\alpha^{-1}\sim N^2.$}
\label{fig:load}
\end{center}
\end{figure}

Figure~\ref{fig:load} shows the results for $N\sum\lambda_\alpha^{-1}$
for the Erd\"os-R\'enyi model as $N$ is increased. Two cases are
considered: when the average nodal degree $d_a$ is 2 and 4. Since
$d_a > 1,$ there is a giant component in each graph, containing an
$N$-independent fraction of the nodes in the large-$N$ limit. All
the other nodes are in components whose size does not diverge as
$N$ is increased. Since we are considering graphs with a single
component in this paper, only the giant component of each graph is
retained. This means that the actual number of nodes in the graph
is a $d_a$-dependent fraction of the $N$ shown in Figure~\ref{fig:load},
but this does not affect the functional form of large-$N$ behavior. 
Each point shown in the figure comes from averaging over eighty random
graphs. We see that $N\sum\lambda_\alpha^{-1}\sim N^2.$

Figure~\ref{fig:load} also shows results for scale free networks.
Following the extension of Ref.~\cite{krapi} of the original
model of Ref.~\cite{barabasi}, nodes enter the network one by one,
with each node born with $p$ edges that link it to pre-existing
nodes; the probability of linking to any preexisting node is
proportional to $d-q$ if its degree is $d,$ where $q$ is a parameter
of the model. The figure shows results for $(p,q) = (2, 0), (3, 0),
(2, 1), (3, 1), (4, 1)$ and $(4, 2).$ As with the Erd\"os-R\'enyi graphs,
each point in the figure comes from averaging over eighty random
graphs. Once again, $N\sum \lambda_\alpha^{-1} \sim N^2.$

\begin{figure}
\begin{center}
\includegraphics[width=3.5in]{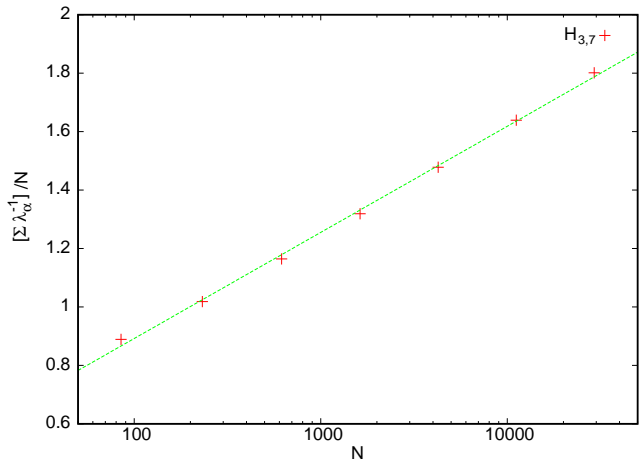}
\caption{Plot of $[\sum_{\alpha\neq 0} \lambda^{-1}_\alpha]/N$
versus $N$ for the hyperbolic grid, with seven triangles meeting at 
every node. All the nodes that are less than some distance $r$ from 
a center node are included; $N$ increases with $r.$ With the $x$-axis 
on a logscale, the straight line fit demonstrates that $N\sum\lambda_\alpha^{-1}\sim N^2\ln N.$}
\label{fig:load_HT}
\end{center}
\end{figure}

Figure~\ref{fig:load_HT} shows the results for $N\sum\lambda_\alpha^{-1}$
for the hyperbolic grid  $\mathbb{H}_{3,7}.$ 
The data
clearly show that $N\sum\lambda_\alpha^{-1} \sim N^2 \ln N.$

\begin{figure}
\begin{center}
\includegraphics[width=3.5in]{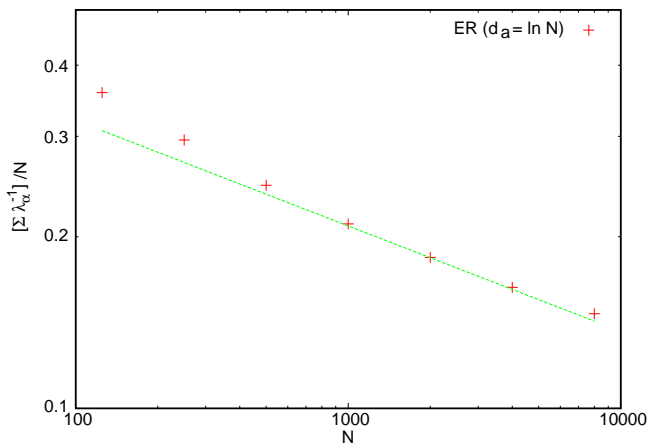}
\caption{Log-log plot of $[\sum_{\alpha\neq 0} \lambda^{-1}_\alpha]/N$
versus $N$ for the Erd\"os-R\'enyi model with the average degree of the nodes
equal to $\ln N.$ The straight line shown corresponds to $0.75 N^{-0.185}.$
}
\label{fig:load_logerdos}
\end{center}
\end{figure}

In the Erd\"os-R\'enyi model, if $d_a = c \ln N$  instead of being
independent of $N,$ there is a phase transition in the behavior of
the model when $c$ is increased to 1: the fraction of the nodes in
the giant component approaches 1. The behavior of graphs constructed
using this model is very different in this regime.
Figure~\ref{fig:load_logerdos} shows the results for
$N\sum\lambda_\alpha^{-1}$ when $d_a = \ln N.$ We see that
$N\sum\lambda_\alpha^{-1}$ grows {\it slower\/} than $N^2$ for large
$N.$ Although the data are not conclusive, they suggest a $\sim
N^{2-\alpha}$ form. As with the other random graph models, each point
in the figure is obtained by averaging over eighty random graphs.

As mentioned earlier in this paper, the maximum load for all the nodes
in a graph consists of --- apart from an additive term --- the
product of $N\sum_\alpha \lambda_\alpha^{-1}$ and the highest nodal
degree in the graph. For scale free graphs, if the probability of
a node having a degree $d$ scales as $p(d)\sim d^{-\gamma}$ for
large $d,$ the highest nodal degree in a graph with $N$ nodes scales
as $N^{1/(\gamma - 1)}$ for large $N.$ 

\section{Conclusions}
We showed for the uniform multicommodity flow problem on an arbitrary
connected graph under random routing, the mean load (or congestion) 
at each node of the graph exists, is unique and derived an explicit 
expression for it in terms of
the spectrum of the graph Laplacian.  Using this explicit expression, 
we obtained analytical
estimates for the mean load for hypercubic lattices and regular trees 
in the large-size regime using their known spectral densities
and computed numerically the mean load
for the Erd\"os-R\'enyi random graphs, the scale-free Barab\'asi-Albert
preferential attachment graphs and hyperbolic grids.

\section{Acknowledgements}

This work of Onuttom Narayan and Iraj Saniee was supported by
grants FA9550-11-1-0278 and 60NANB10D128 from AFOSR and NIST. 


{}
\end{document}